\documentclass[aps,twocolumn,showpacs,preprintnumbers,nofootinbib,superscriptaddress]{revtex4}
\usepackage{amsmath} \usepackage{graphicx} \usepackage{amsfonts}
\usepackage{array} \usepackage{amsthm} \usepackage{bm} 

\usepackage[breaklinks]{hyperref}
\usepackage{color}

\newcommand{\nn}{\nonumber}
\newcommand{\be}{\begin{equation}}
\newcommand{\ee}{\end{equation}}
\newcommand{\ba}{\begin{eqnarray}}
\newcommand{\ea}{\end{eqnarray}}
\newcommand{\bal}{\begin{align}}
\newcommand{\eal}{\end{align}}

\newcommand{\e}{{\rm e}}
\newcommand{\dd}{{\rm d}}

\newcommand{\bb}{\bibitem}

\newcommand{\om}{\omega}
\newcommand{\al}{\alpha}
\newcommand{\la}{\lambda}
\newcommand{\La}{\Lambda}
\newcommand{\bt}{\beta}

\newcommand{\ga}{\gamma}

\newcommand{\ep}{\epsilon}

\newcommand{\ta}{\theta}
\newcommand{\Ta}{\Theta}

\newcommand{\De}{\Delta}

\newcommand{\de}{\delta}

\newcommand{\bw}{\begin{widetext}}
\newcommand{\ew}{\end{widetext}}

\def\reh{r_{\text{eh}}}

\begin{document}
\title{Black hole thermodynamics: No inconsistency via the inclusion of the missing P-V terms}

\author{Mustapha Azreg-A\"{\i}nou}
\affiliation{Ba\c{s}kent University, Faculty of Engineering, Ba\u{g}l\i ca Campus, 06810 Ankara, Turkey}


\begin{abstract}
The early literature on black hole thermodynamics ignored the $P$-$V$ term associated with the existence of a fundamental physical constant in the black hole solution. The inclusion of this constant in the first law becomes inconsistent with the Smarr relation. Once the missing $P$-$V$ term introduced is, it becomes customary to introduce it only in problems where there is a negative cosmological constant. This practice is inherited from cosmological approaches which consider the quantity $-\Lambda/8\pi$ as the constant pressure due to a cosmological fluid. However, the notions of pressure and thermodynamic volume in black hole thermodynamics are very different from their counterparts in classical thermodynamics. From this point of view, there is \textit{a priori} no compelling reason to not extend this notion of pressure and associate a partial pressure with each ``external" density $8\pi T_{t}{}^{t}$. In this work, we associate a partial pressure with a variable mass parameter as well as with each $tt$ component of the effective stress-energy tensor $T_{\text{eff}\,\mu}{}^{\nu}$ but not with the linear component of the electromagnetic field. Using the field equations $G_{\mu}{}^{\nu}=8\pi T_{\text{eff}\,\mu}{}^{\nu}$, we derive universal expressions for the enthalpy, internal energy, free energies, thermodynamic volume, equation of state, law of corresponding states, criticality, and critical exponents of static (nonrotating) charged black holes, with possibly a variable mass parameter, whether they are solutions to the Einstein field equations or not. We extend the derivation to the case where the black hole is immersed in the field of a quintessence force and to the multiforce case. Many applications and extensions are considered, including applications to regular black holes derived in previous and present work. No inconsistency has been noticed in their thermodynamics.
\end{abstract}


\maketitle

\section{Introduction\label{secI}}

For consistency in the thermodynamic laws, it has become customary to treat (some of) the physical constants as thermodynamic variables~\cite{ex0}-\cite{AzregE}. If the conjugate variable associated with the physical constant $C$ is denoted by $\Ta$, the corresponding term $\Ta \dd C$, added to the first law of thermodynamics, generally lacks a direct physical meaning~\cite{AzregE}. It is preferable to introduce a new thermodynamic variable and its conjugate which both have a familiar physical meaning. These variables make up the missing $P$-$V$ term in the first law of thermodynamics. Such a treatment should apply to any fundamental theory with many physical constants~\cite{iso,AzregE,consts}.

Such a $P$-$V$ term has arisen in all theories of gravitation where an extra negative cosmological density is taken into consideration. Very recently we have associated a $P$-$V$ term with both a positive cosmological density and a quintessential one~\cite{AzregE}. Since a quintessential density is not constant, the associated pressure is the quintessential pressure evaluated at the horizon, and $V$ is its corresponding conjugate thermodynamic volume with respect to the enthalpy of the hole.

We aim to generalize this to any density included in the rhs of the field equations $G_{\mu}{}^{\nu}=8\pi T_{\mu}{}^{\nu}$ (we use geometric units $G=\hbar=c=k_{\text{B}}=1$). In this approach, however, to have criticality, we associate no pressure ($P=0$) with a linear electromagnetic density; that is, we treat an electric charge of a linear electromagnetic density as an intrinsic property of the black hole rather than an external force. Said otherwise, we dispose of a linear electromagnetic density as an internal part of the thermodynamic system. If a nonlinear electromagnetic density is present in the rhs of the field equations, as is the case with the Born-Infeld theory~\cite{BI}, this will be split into a linear part with no pressure associated with, and a pressure associated with the remainder of the density.

This generalization extends to include the black holes of modified theories of general relativity, as Born-Infeld theory. Their field equations can always be brought to the form $G_{\mu}{}^{\nu}=8\pi T_{\text{eff}\,\mu}{}^{\nu}$, where $T_{\text{eff}\,\mu}{}^{\nu}$ is the effective stress-energy tensor (SET) and $8\pi T_{\text{eff}\,t}{}^{t}$ is the effective external density.

As we shall see, the effect of the energy densities is linear, in that each added contribution to the rhs of the field equations results in a $P$-$V$ term in the first law of black hole thermodynamics.

We shall also show that the notions, and the final expressions, of the thermodynamic potentials and thermodynamic volume depend on how one disposes of the densities as internal or external forces.

Let the metric of the static spacetimes be of the form
\begin{equation}\label{1.1}
    \dd s^2=f(r)\dd t^2-f^{-1}(r)\dd r^2-r^2\dd \Omega^2
\end{equation}
with signature ($+,-,-,-$). We assume that $f(r)=0$ has at least one root; in case $f(r)=0$ has many roots, we will be concerned with the nonextremal event horizon $\reh$, which is the largest root satisfying
\begin{equation}\label{1.2}
    f(\reh)=0 \;\text{ and }\; f_{,r}(\reh)>0.
\end{equation}
Throughout this note, we use the indexical and usual notations for derivatives: $f_{,x}\equiv \partial f/\partial x$. At the other roots of $f(r)=0$, if there are any that are larger than $\reh$, the derivative is negative.

We assume that the metric~\eqref{1.1} describes a static (nonrotating) charged black hole with a variable mass parameter $M\equiv M(r)$. Such a metric describes all types of black holes including regular cosmological black holes with a de Sitter core~\cite{irina}. Throughout this work, we dispose of the electric charge of a linear electromagnetic density as an internal, intrinsic property of the black hole; that is, the energy density due to the electric field is not treated as an external, extra density added to the rhs of the field equations. In Sec.~\ref{secE}, however, we will deal with the case where we dispose of the electric energy density as an external force. With that said, the general expression of $f$ reads
\begin{equation}\label{1.3}
    f(r)=1-\frac{2M(r)}{r}+\frac{Q^2}{r^2}-g(r,C)
\end{equation}
where $C$ is some physical constant, which we will omit in most cases. Note that $g$ may depend also on $Q$ as well as on $M$, which we will omit too.

The assumption $M\equiv M(r)$ allows for a general treatment including fluid solutions used as regular cores~\cite{irina}-\cite{conf}. In cases where $\lim_{r\to\infty}M(r)\equiv M_{\infty}$ exists, it is equivalent, in the context of our approach, to treat the mass parameter as constant by writing
\begin{equation*}
\frac{2M(r)}{r}=\frac{2M_{\infty}}{r}+\Big(\frac{2M(r)}{r}-\frac{2M_{\infty}}{r}\Big),
\end{equation*}
and including the expression inside the parentheses into $g(r)$ [an instance of this application is provided in Sec.~\ref{secfa4}].

In Sec.~\ref{secnq} we consider the case with one physical constant then generalize to the multi-physical-constant black hole. We determine all the relevant thermodynamic potentials, thermodynamic volume, equation of state (EOS), and the expression for the pressure. In Sec.~\ref{secfg} we generalize the results to the case where the area law for the entropy no longer holds. Section~\ref{seca} is split into two parts: in the first one we discuss the criticality conditions for all black holes with special application to AdS ones, and in the second one we calculate the critical exponents for black holes obeying the area law for entropy.

Other applications are given in Sec.~\ref{secfa} where we apply the results to Born-Infeld-AdS, Born-Infeld-dS, Martinez-Troncoso-Zanelli~\cite{MTZ}, and three regular black holes one of which is determined in this work.

In Sec.~\ref{secq} we further generalize the results to include multi-physical-constant black holes surrounded by quintessence and provide an application. Finally, the question is discussed of how the $P$-$V$ term could be introduced in the early literature on black hole thermodynamics with two applications given in Sec.~\ref{secE}. We conclude in Sec.~\ref{secc}. Two Appendixes have been added: Appendix A shows the relation between the enthalpy and the mass parameter, and Appendix B derives the differential equation to which the thermodynamic volume is a solution for black holes not obeying the area law for entropy.

\section{Thermodynamic potentials---no quintessence \label{secnq}}

From now on, we use the new radial variable,
\begin{equation}\label{nq1}
    u=r^2.
\end{equation}
The entropy of the event horizon is assumed to be proportional to the event horizon area
\begin{equation}\label{nq2}
    S=\pi \reh{}^2=\pi s,
\end{equation}
where, almost throughout this paper, we will use conveniently the variable
\begin{equation}\label{nq2b}
s\equiv \reh{}^2
\end{equation}
instead of the horizon radius $\reh$.

We introduce the one-variable function $E(u)$ defined by
\begin{equation}\label{nq3}
    E(u)\equiv \frac{\sqrt{u}}{2}+\frac{Q^2}{2\sqrt{u}}-\frac{\sqrt{u}}{2}g(u,C),
\end{equation}
in terms of which we reexpress $f(u)$ as
\begin{equation}\label{nq4}
    f(u)=\frac{2}{\sqrt{u}} [E(u)-M(u)].
\end{equation}
On the nonextremal event horizon, we have $f(s)=0$, that is, $E(s)=M(s)$ but $E_{,s}(s)\neq M_{,s}(s)$; rather, we have $E_{,s}(s)> M_{,s}(s)$ since the temperature of the nonextremal event horizon, defined by $T=f_{,r}/(4\pi)|_{r=\reh}$, is positive as prescribed in~\eqref{1.2},
\begin{align}
\label{nq5}T=&\frac{\sqrt{u}}{2\pi}f_{,u}\Big|_{u=s}=\frac{E_{,s}(s)- M_{,s}(s)}{\pi}>0,\\
\label{nq5b}\quad =&\frac{1}{4\pi \sqrt{s}}-\frac{Q^2}{4\pi s^{3/2}}-\frac{(\sqrt{s}g+2M)_{,s}}{2\pi},
\end{align}
where we have used $E(s)=M(s)$ in the first line. It is understood that the expressions $E_{,s}(s)- M_{,s}(s)$ and $(\sqrt{s}g+2M)_{,s}$ are obtained upon first taking the derivative with respect to $u$ then evaluating at $u=s$. On the event horizon, the other important thermodynamic variable is the electric potential
\begin{equation}\label{nq5c}
    A=\frac{Q}{\sqrt{s}}=\frac{\sqrt{\pi}Q}{\sqrt{S}}.
\end{equation}

If $\ep$ denotes the total energy, $\ep\equiv 8\pi T_{t}{}^{t}$, then the field equation $G_{t}{}^{t}=8\pi T_{t}{}^{t}$ yields
\begin{equation}\label{nq6}
2(\sqrt{u}g+2M)_{,u}=\sqrt{u}\ep_{\text{ext}},
\end{equation}
where we have set
\begin{equation}\label{nq7}
    \ep_{\text{ext}}\equiv \ep - \frac{Q^2}{u^2},
\end{equation}
which is the extra added density, that is, the total density $\ep$ reduced by the linear electric energy density $Q^2/u^2$.

Note that our assumptions $g_{tt}=-1/g_{rr}$ and $g_{\ta\ta}=g_{\varphi\varphi}/\sin^2\ta=-r^2$ [Eq.~\eqref{1.1}] imply that $G_{t}{}^{t}=G_{r}{}^{r}$. If the metric~\eqref{1.1} is solution to $G_{\mu}{}^{\nu}=8\pi T_{\mu}{}^{\nu}$, this results in $T_{t}{}^{t}=T_{r}{}^{r}$ for the on-shell values of the metric and the fields defining the SET. Recall that only the on-shell values are needed for the thermodynamic description of black holes. The on-shell relation $T_{t}{}^{t}=T_{r}{}^{r}=\ep/8\pi$ yields the following common and familiar definition of the radial pressure $P$ exerted on the event horizon\footnote{In the presence of quintessence, we assign an average value over the angles to the spatial components of the SET which is proportional to $T_{t}{}^{t}$~\cite{Kis,AAR}. So, in all cases where $g_{tt}=-1/g_{rr}$ and $g_{\ta\ta}=g_{\varphi\varphi}/\sin^2\ta=-r^2$, $T_{t}{}^{t}$ is convenient for defining the pressure on the horizon.}:
\begin{equation}\label{nq8}
    P=-\frac{\ep_{\text{ext}}}{8\pi}\Big|_{u=s}=-\frac{\ep_{\text{ext}}(s)}{8\pi}.
\end{equation}
This does not include the pressure $-Q^2/(8\pi u^2)$ due to the linear part of the electric field density, which has been removed because it is treated as an intrinsic property (see Sec.~\ref{secE} for comments); however, it includes the pressure $-M(s)_{,s}/(2\pi\sqrt{s})$ due to the mass gradient. Thus, the latter pressure is considered as an external effect. The other two nonradial components of the pressure are irrelevant for the thermodynamic of spherically symmetric black holes. Using~\eqref{nq6}, $P$ reads
\begin{equation}\label{nq9}
    P=-\frac{[\sqrt{s}g(s,C)+2M(s)]_{,s}}{4\pi \sqrt{s}}.
\end{equation}

We search for a function $H(S,Q,P)$ satisfying the properties,
\begin{align}
\label{nq10a}&\Big(\frac{\partial H}{\partial S}\Big)_{Q,P}=T, \\
\label{nq10b}&\Big(\frac{\partial H}{\partial Q}\Big)_{S,P}=A ,
\end{align}
where $T$ and $A$ are given in~\eqref{nq5b} and~\eqref{nq5c}, respectively. Here $(S,Q,P)$ are considered as independent thermodynamic variables; this is possible since $P$ depends on ($S,C$), so it is possible to vary $S$ and $P$ independently, recalling that $C$ is considered as a thermodynamic variable. If $g$ does not depend on any physical constant $C$, as in the examples treated in Sec.~\ref{secfa4} where the mass parameter is constant, we use the latter as an independent thermodynamic variable or any other parameter other than the electric charge, and in this case $P$ depends on $S$ and on the other parameter.

By a similar argument used in Ref.~\cite{AzregE}, it is straightforward to show that $E(s)$ is not the appropriate function $H$, for it does not satisfy the requirement~\eqref{nq10a}. Following Ref.~\cite{AzregE}, we look for the expression of $H(S,Q,P)$ in the form
\begin{equation}\label{nq12}
\frac{\sqrt{s}}{2}+\frac{Q^2}{2\sqrt{s}}+V(s)P=\frac{\sqrt{S}}{2\sqrt{\pi}}+\frac{\sqrt{\pi}Q^2}{2\sqrt{S}}+V(S)P.
\end{equation}
Here $V(s)$ is the so-called thermodynamic volume to be fixed by the requirement~\eqref{nq10a}. Using~\eqref{nq5b} in~\eqref{nq10a}, we arrive at
\begin{equation}\label{nq13}
\frac{V_{,s}P}{\pi}=
-\frac{(\sqrt{s}g+2M)_{,s}}{2\pi},
\end{equation}
which is a differential equation. Using~\eqref{nq9} to eliminate $P$, we arrive at the simplified differential equation
\begin{equation}\label{nq14}
V_{,s}=2\pi \sqrt{s},
\end{equation}
the solution of which is given by
\begin{equation}\label{nq15}
    V(s)=\frac{4\pi}{3}s^{3/2}.
\end{equation}
The final expression of $H$ reads along with the necessary formulas:
\begin{align}
&H(S,Q,P)=\frac{\sqrt{s}}{2}+\frac{Q^2}{2\sqrt{s}}+\frac{4\pi}{3}s^{3/2}P,\nn\\
&P=-\frac{[\sqrt{s}g(s,C)+2M(s)]_{,s}}{4\pi \sqrt{s}},\nn\\
\label{nq16}&V\equiv \Big(\frac{\partial H}{\partial P}\Big)_{S,Q}=\frac{4\pi}{3}s^{3/2},\\
&T=\frac{1}{4\pi \sqrt{s}}-\frac{Q^2}{4\pi s^{3/2}}+2\sqrt{s}P,\nn\\
&\dd H = T \dd S+A\dd Q+V\dd P.\nn
\end{align}
Note that the thermodynamic volume $V$ is equal to the geometric volume. The last line in~\eqref{nq16}, where the differential of $H(S,Q,P)$ is expressed in terms of the thermodynamic variables ($T,A,V$) as in classical thermodynamics, allows us to interpret $H$ as the enthalpy of the black hole.

For a Schwarzschild black hole, the enthalpy $H$ and the internal energy $U=H-PV$ reduce to $\sqrt{s}/2=M$. It is straightforward to check that if the mass parameter is constant, then $H=M$ for a Schwarzschild de Sitter (dS) or anti--de Sitter (AdS) black hole. For a general black hole solution, we explore the relation between the enthalpy and the constant or variable mass parameter in Appendix A. We will particularly reach the conclusion that the pressure~\eqref{nq9} for any black hole solution, where $H=M$, reduces to a simple expression $P=-3g(s)/(8\pi s)$ generalizing that of the charged dS and AdS black holes, which is $P_{\La}=-\La/(8\pi)$ with $g_{\La}(s)=\La s/3$. Another conclusion that we will draw in Appendix A is the discrepancy between the enthalpy and the mass parameter for the dS solutions with variable mass parameter used as regular cores~\cite{irina}.

It is worth noticing that the expression of the enthalpy given in~\eqref{nq16} is not the unique solution to the system of differential equations~\eqref{nq10a} and~\eqref{nq10b}. However, this expression is the physical one satisfying all the requirements of classical thermodynamics. For instance, one may derive another, but nonphysical, expression for $H$, as the following one, satisfying the requirements~\eqref{nq10a} and~\eqref{nq10b},
\begin{multline}\label{nq16b}
\bar{H}=\frac{\sqrt{s}}{2}+\frac{Q^2}{2\sqrt{s}}-\al M(s)\\
 +\frac{4\pi}{3}\Big[s^{3/2}-3\al \int^s \frac{\sqrt{s'}M(s')_{,s'}}{[\sqrt{s'}g(s')+2M(s')]_{,s'}}\,\dd s'\Big]P,
\end{multline}
where the thermodynamic volume $\bar{V}$ is the coefficient of $P$ in~\eqref{nq16b} and $\al$ is any constant, but the corresponding internal energy does not reduce to $M$ for a Schwarzschild black hole. Moreover, the internal energy $\bar{H}-P\bar{V}$ for any black hole becomes a function of the arbitrary constant $\al$.

\subsection*{Multi-physical-constant case}

The generalization to the multi-physical-constant case is straightforward. If $f$ is of the form
\begin{equation}\label{g1}
    f(r)=1-\frac{2M(r)}{r}+\frac{Q^2}{r^2}-\sum_i g_i(r,C_i),
\end{equation}
using the fact that the differential equation~\eqref{nq6} is linear, we obtain the generalized formulas:
\begin{align}
&H(S,Q,P)=\frac{\sqrt{s}}{2}+\frac{Q^2}{2\sqrt{s}}+\frac{4\pi}{3}s^{3/2}P,\nn\\
&P=\sum_i P_i-\frac{M(s)_{,s}}{2\pi\sqrt{s}},\quad P_i=-\frac{[\sqrt{s}g_i(s,C_i)]_{,s}}{4\pi \sqrt{s}},\nn\\
\label{g2}&V\equiv \Big(\frac{\partial H}{\partial P}\Big)_{S,Q}=\frac{4\pi}{3}s^{3/2},\\
&T=\frac{1}{4\pi \sqrt{s}}-\frac{Q^2}{4\pi s^{3/2}}+2\sqrt{s}P,\nn\\
&\dd H = T \dd S+A\dd Q+V\dd P. \nn
\end{align}
It is worth noticing that the first two terms in the expression of $T$ are the contributions of the Reissner-Nordstr\"om event horizon and that each external density, including the mass gradient, contributes additively to the temperature of the black hole; positive densities reduce the temperature of the hole and negative densities have the opposite effect.

Now, if we reexpress $H$~\eqref{g2} as a function of ($S,Q^2,P^{-1}$),
\begin{equation*}
H(S,Q^2,P^{-1})=\frac{\sqrt{s}}{2}+\frac{Q^2}{2\sqrt{s}}+\frac{4\pi}{3}s^{3/2}(P^{-1})^{-1},
\end{equation*}
we see that
\begin{equation*}
H(\la S,\la Q^2,\la P^{-1})=\la^{1/2}H(S,Q^2,P^{-1}),
\end{equation*}
which shows that $H$ is homogeneous in ($S,Q^2,P^{-1}$) of order $1/2$, and the Euler identity for thermodynamic potentials that are not homogeneous functions of their natural extensive variables~\cite{on-gtd} yields the Smarr formula
\begin{equation}\label{g3}
  H=2T S+AQ-2VP.
\end{equation}

The internal energy $U(S,Q,V)=H-PV$ has the following universal expression for all black holes:
\begin{equation}
\label{g3b}U=\frac{\sqrt{s}}{2}+\frac{Q^2}{2\sqrt{s}}=\frac{\sqrt{S}}{2\sqrt{\pi}}+\frac{\sqrt{\pi}Q^2}{2\sqrt{S}},
\end{equation}
which depends only on the intrinsic properties of the black hole and does not depend explicitly on the external densities; this corresponds well to the definition of the internal energy in classical thermodynamics. Rather, $U$ depends on the external densities only implicitly through $s$ which is a root of $f(s)=0$. $V$ depends explicitly only on $S$, and so it is a redundant variable as in the $M=\text{const}$ case~\cite{Dolan,AzregE}.

In the remaining part of this section, we aim to derive general expressions for the other thermodynamic potentials, mainly, the Helmholtz $F$ and Gibbs $G$ free energies and the critical exponents.\\

\paragraph*{\textbf{Thermodynamic potentials.}}

Using~\eqref{g3}, $U=H-PV$, $F=U-TS$, and $G=H-TS$ we obtain
\begin{equation}\label{s1}
F=T S+AQ-3VP\quad \text{and}\quad G=T S+AQ-2VP.
\end{equation}
These expressions yield
\begin{align}
F(T,V,Q)&=T\pi s+\frac{Q}{\sqrt{s}}Q-4\pi s^{3/2}\Big[\frac{T}{2\sqrt{s}}-\frac{1}{8\pi s}+\frac{Q^2}{8\pi s^2}\Big]\nn\\
\label{s2}\quad &=-\pi Ts+\frac{\sqrt{s}}{2}+\frac{Q^2}{\sqrt{s}}
\end{align}
for the Helmholtz free energy, where we have used the fourth line in~\eqref{g2}, and
\begin{align}
G(T,P,Q)&=\Big[\frac{1}{4\pi \sqrt{s}}-\frac{Q^2}{4\pi s^{3/2}}+2\sqrt{s}P\Big]\pi s+\frac{Q}{\sqrt{s}}Q\nn\\
\quad &\quad -\frac{8\pi s^{3/2}}{3}P\nn\\
\label{s3}\quad &=\frac{\sqrt{s}}{4}-\frac{2\pi s^{3/2}}{3}P+\frac{3Q^2}{4\sqrt{s}}
\end{align}
for the Gibbs free energy, where we have used the fourth line in~\eqref{g2}. In~\eqref{s2}, $s$ is a function of $V$: $s=[3V/(4\pi)]^{2/3}$, while in~\eqref{s3}, $s$ is an implicit solution to the fourth line in~\eqref{g2} (the EOS) expressed in terms of ($T,P,Q$).

The expressions~\eqref{s2} and~\eqref{s3} apply to all black holes if $S=\pi s$ is the law for the entropy. They have been derived in Ref.~\cite{Mann1}, by other procedures, for the special case of the charged AdS black holes [Eqs.~(3.29) and~(3.28) of Ref.~\cite{Mann1}, respectively].

\section{Further generalization: $\pmb{S\neq \pi \reh{}^2}$\label{secfg}}

There have been a couple of entropy formulas modifying the usual area one, $S=\pi s$. These modifications are either corrections to the area law~\cite{cor1}-\cite{cor4} or general prescriptions to identify the entropy of a stationary black hole in any spacetime dimension as a Noether charge~\cite{extra1}-\cite{extra3}. To the best of our knowledge, these modifications include the physical constants $C_i$ and the electric charge $Q$ and so on. Whatever the type of the modification made to the expression of the entropy, we can assume $S$ of the form [we keep using conveniently the variable $s\equiv \reh{}^2$ introduced in~\eqref{nq2b}]
\begin{equation}\label{g4}
    S=\pi h(s,C_i,Q),
\end{equation}
where $h(s)\equiv s$ if the area law applies.

If $S$ depends explicitly only on ($s,C_i$), then the enthalpy and internal energy are still given by the r.h.s.'s of~\eqref{nq12} and~\eqref{g3b}, respectively, with $S=\pi h(s,C_i)$,
\begin{align}
\label{g4b}&H=\frac{\sqrt{S}}{2\sqrt{\pi}}+\frac{\sqrt{\pi}Q^2}{2\sqrt{S}}+V(S,C_i)P,\\
\label{g4c}&U=\frac{\sqrt{S}}{2\sqrt{\pi}}+\frac{\sqrt{\pi}Q^2}{2\sqrt{S}},\quad [S=\pi h(s,C_i)].
\end{align}
$V$ is now a function of the $C_i$'s and its determination is no longer straightforward. Even in the simplest case where $S$ depends only on one physical constant $C$ and on $s$, Eq.~\eqref{nq13} takes rather a more complicated shape which is generally not tractable analytically (see Appendix B). We will not pursue this program any further. It is worth mentioning, however, that in this case, since $V$ depends on ($s,C$), it is possible to vary independently $S$ and $V$, and so $V$ is no longer a redundant variable.

Another point worth mentioning is that the fourth equation in~\eqref{g2} remains valid, too, in the general case where~\eqref{g4} applies. This is the EOS rewritten as
\begin{equation}\label{g5}
P=\frac{T}{2\sqrt{s}}-\frac{1}{8\pi s}+\frac{Q^2}{8\pi s^2}.
\end{equation}

The approach developed in this section has led, in the case where the area law for entropy applies, to universal formulas for the enthalpy, internal energy, thermodynamic volume, and EOS. In the case where the area law for entropy is altered, the only remaining valid formula\footnote{If $S$ is independent of $Q$, the expressions of the enthalpy and internal energy are given by~\eqref{g4b} and~\eqref{g4c}. $V$ is determined by the requirement~\eqref{nq10a} as done in Appendix B.} is that of the EOS~\eqref{g5}.

Whether the area law for entropy applies or not, the approach developed in this section applies to all black holes, even to black holes that are solutions to the field equations of modified theories of general relativity, such as the Born-Infeld theory, provided $g_{tt}=-1/g_{rr}$ and $g_{\ta\ta}=g_{\varphi\varphi}/\sin^2\ta=-r^2$. The field equations can always be brought to the form $G_{\mu}{}^{\nu}=8\pi T_{\text{eff}\,\mu}{}^{\nu}$, where $T_{\text{eff}\,\mu}{}^{\nu}$ is the effective SET. In this case, the expressions~\eqref{nq6} of $\ep_{\text{ext}}$ and~\eqref{nq9} of $P$ are definitions of the \textit{effective} extra densities and \textit{effective} \textit{total} pressure. Here each density contributes a partial pressure, and the total pressure is due to all densities. Even a variation in the mass parameter contributes a partial pressure; only densities due to linear electrodynamics do not contribute to the total pressure (see Sec.~\ref{secE} for an alternative discussion).

Contrary to what one finds in the literature, where only the negative cosmological constant density generates the total pressure, our thermodynamic ensemble is such that the total pressure is generated by all densities except the linear-electrodynamic density. If a nonlinear-electrodynamic density is present, this is split into a linear part, which does not contribute to the total pressure, and a remainder part contributing to the total pressure. The advantage of doing so is to have a universal EOS~\eqref{g5}, universal expressions of the thermodynamic potentials and criticality, and universal critical exponents for all black holes.

Moreover, our thermodynamic ensemble introduces a new thermodynamic concept, that is, the concept of the ``effective thermodynamic electric or magnetic potential" similar to that of the thermodynamic volume (generally different from the geometric volume) which is, in general, different from the physical potential. If $S$ depends explicitly only on ($s,C_i$), then the thermodynamic electric potential is given by the second expression in the rhs of~\eqref{nq5c} with $S=\pi h(s,C_i)$
\begin{equation}\label{g6}
\Phi=\frac{\sqrt{\pi}Q}{\sqrt{S}},\quad [S=\pi h(s,C_i)].
\end{equation}
With this expression of $A$ and that of $H$, given by~\eqref{g4b}, the identity~\eqref{g3} still holds if $S=\pi h(s,C_i)$.

It is worth emphasizing that, in the derivation of the EOS~\eqref{g5}, all we have employed is the notion of the effective total pressure; no notion of the concept of the effective thermodynamic potential was used. This is to say that the value of the last term in~\eqref{g5}, which is the contribution of the linear part of the (nonlinear, if any) electromagnetic density, is invariant and unchanged under the transformation $s\to h$, while the value of the thermodynamic potential is altered.

\section{Applications \label{seca}}

\subsection{General considerations}

Following Ref.~\cite{Mann1} we identify the specific volume $v$ by introducing the physical thermodynamic quantities $\text{Pressure}\equiv \hbar cP/\ell_{\text{P}}{}^2$ and $\text{Temperature}\equiv \hbar cT/k_{\text{B}}$, where $\ell_{\text{P}}=\sqrt{\hbar G/c^3}$ is the Planck length. This has led to
\begin{equation}\label{a1}
    v=2\ell_{\text{P}}{}^2\reh \quad (=2\ell_{\text{P}}{}^2\sqrt{s}).
\end{equation}
Using this in the EOS~\eqref{g5}, this reduces to
\begin{equation}\label{eos}
P=\frac{T}{v}-\frac{1}{2\pi v^2}+\frac{2Q^2}{\pi v^4},
\end{equation}
where we use again the geometric units in which $v=2\reh$. This is precisely Eq.~(3.15) of Ref.~\cite{Mann1} which was derived for the charged AdS black hole, where $P$ denoted the pressure due to a negative cosmological constant: $P=-\La/8\pi$. In our Eq.~\eqref{eos}, $P$ denotes the effective total pressure due to all densities but the linear part of the electrodynamic density.

So most of the results derived in Sec. III of Ref.~\cite{Mann1} apply to our general case~\eqref{eos}. There is a critical temperature $T_c$ below which a $P-v$ diagram of isotherms shows a point of inflection as in a $P-v$ diagram of the Van der Waals gas-liquid system. $T_c$ and the corresponding critical volume $v_c$ and pressure $P_c$ are solutions to
\begin{equation}\label{eosd}
    \Big(\frac{\partial P}{\partial v}\Big)_{T,Q}=0 \;\text{ and }\;\Big(\frac{\partial^2 P}{\partial v^2}\Big)_{T,Q}=0,
\end{equation}
which yield the values
\begin{equation}\label{a2}
T_c=\frac{1}{3\sqrt{6}\pi |Q|},\; v_c=2\sqrt{6} |Q|,\; P_c=\frac{1}{96\pi Q^2}.
\end{equation}
preserving the value 3/8 for the ratio $P_cv_c/T_c$, known for the Van der Waals gas-liquid system, but not the shape of the law of corresponding states which reads
\begin{equation}\label{a2a}
8t=3\nu\Big(p+\frac{2}{\nu^2}\Big)-\frac{1}{\nu^3},
\end{equation}
where $T=T_ct$, $v=v_c\nu$, and $P=P_cp$. This is different from the law of corresponding states for the Van der Waals gas-liquid system.

Another interesting point is where the isotherm becomes tangent to the $v$ axis in a $P-v$ diagram. This is the point
\begin{equation}\label{a2b}
T_0=\frac{1}{6\sqrt{3}\pi |Q|},\;v_0=2\sqrt{3} |Q|,\;P_0=0,
\end{equation}
where the rhs of~\eqref{eos} has a double positive root ($2\sqrt{3} |Q|$) and a negative one ($-\sqrt{3} |Q|$). Notice the independence on the charge $Q$ of the following quantities:
\begin{equation*}
2T_0v_0=T_cv_c=\frac{2}{3\pi},\;\frac{T_c}{T_0}=\frac{v_c}{v_0}=\sqrt{2},
\end{equation*}
which, however, do possess a universal character only for black holes and they are not satisfied by the corresponding values of the Van der Waals gas-liquid system.

Thus, we have thus shown, by a judicious choice of thermodynamic variables ($P,v,T$), the possible existence, and universality, of criticality for the fixed $Q$ subensemble for all black holes, whatever the relation between the specific volume $v$ (the radius of the event horizon $\reh$) and the thermodynamic volume $V$ is.

The existence of criticality depends on the sign of the pressure~\eqref{nq9}. For the Reissner-Nordstr\"om AdS black hole, $P=P_{\La}\equiv -\La/8\pi>0$, but this does not guarantee the existence of criticality for any value of $Q$ unless $P_c<P_{\La}$, yielding
\begin{equation}\label{a2bb}
    Q^2>-1/(12\La).
\end{equation}

A Reissner-Nordstr\"om black hole exists only if $Q^2\leq M^2$; thus, we have reached the following constraint on the existence of criticality for the charged AdS black hole:
\begin{equation}\label{a2c}
    M>M_c\equiv \sqrt{-\frac{1}{12\La}}.
\end{equation}
Moreover, since $\La$ is a thermodynamic variable in the context of modern black hole thermodynamics, fluctuations in its value $\De\La$ are constrained by $|\De\La|\ll|\La|$ to ensure positiveness of the pressure; therefore, $T_0$~\eqref{a2b} is a lower-limit temperature for the charged AdS black holes. $\La$ being a small number, by~\eqref{a2c} we see that only huge black holes, if any, may undergo criticality. It is easy to see that this critical behavior does not take place near extremality, as this can be shown on substituting the value of $r_c=v_c/2=\sqrt{6} |Q|$ into $f(r)=0$, yielding the constraint
\begin{equation}\label{a2d}
    \frac{M}{|Q|}= \frac{(7-12\La Q^2)}{2\sqrt{6}}>\frac{4}{\sqrt{6}}\simeq 1.6,
\end{equation}
where we have used~\eqref{a2bb}.

\begin{figure}
\centering
  \includegraphics[width=0.40\textwidth]{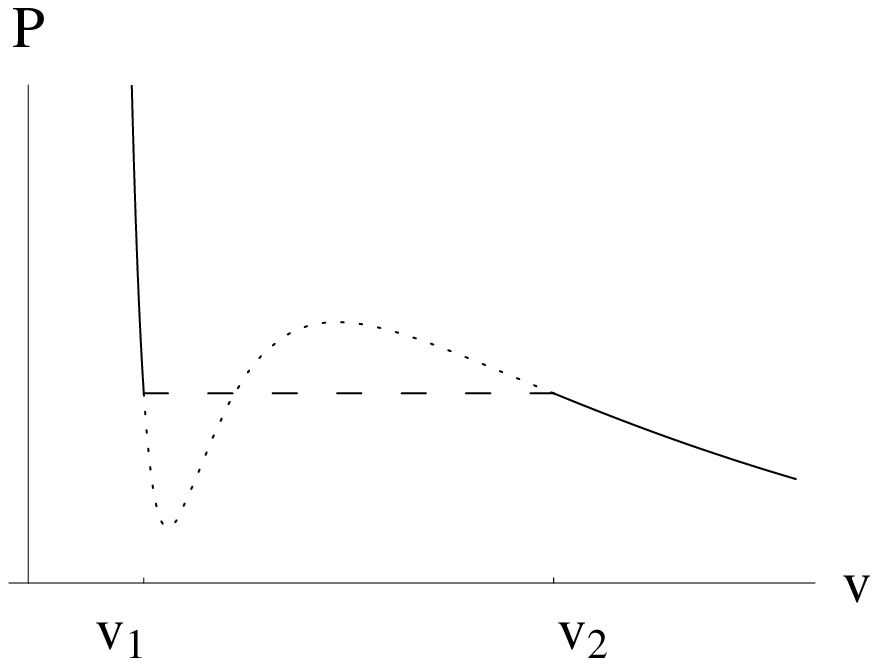}\\
  \caption{\footnotesize{A generic $P-v$ diagram showing an isotherm for $T_0<T<T_c=1/(3\sqrt{6}\pi |Q|)$ and an isobar $P=\text{const}>0$ (dashed line). The Maxwell's equal area law applies to the oscillating part of the isotherm $T_0<T<T_c$ (dotted curve). This is expressed by the line integral $\oint v\,\dd P=0$.}}\label{Fig1}
\end{figure}

For black holes where the pressure~\eqref{nq9} is positive, as is the case for the Reissner-Nordstr\"om AdS one, there are isotherms with an oscillating part that is totally above the $v$ axis ($T_0<T<T_c$), as shown in the generic Fig.~\ref{Fig1}, and other isotherms with an oscillating part only partly above the $v$ axis ($0<T<T_0$). The transition from the small black hole to the large one may proceed along the isobar $P=\text{const}>0$, as shown in Fig.~\ref{Fig1}, where the value of $P$ is selected on observing Maxwell's equal area law,
\begin{equation*}
    \oint v\,\dd P=0.
\end{equation*}

\begin{figure}
\centering
  \includegraphics[width=0.40\textwidth]{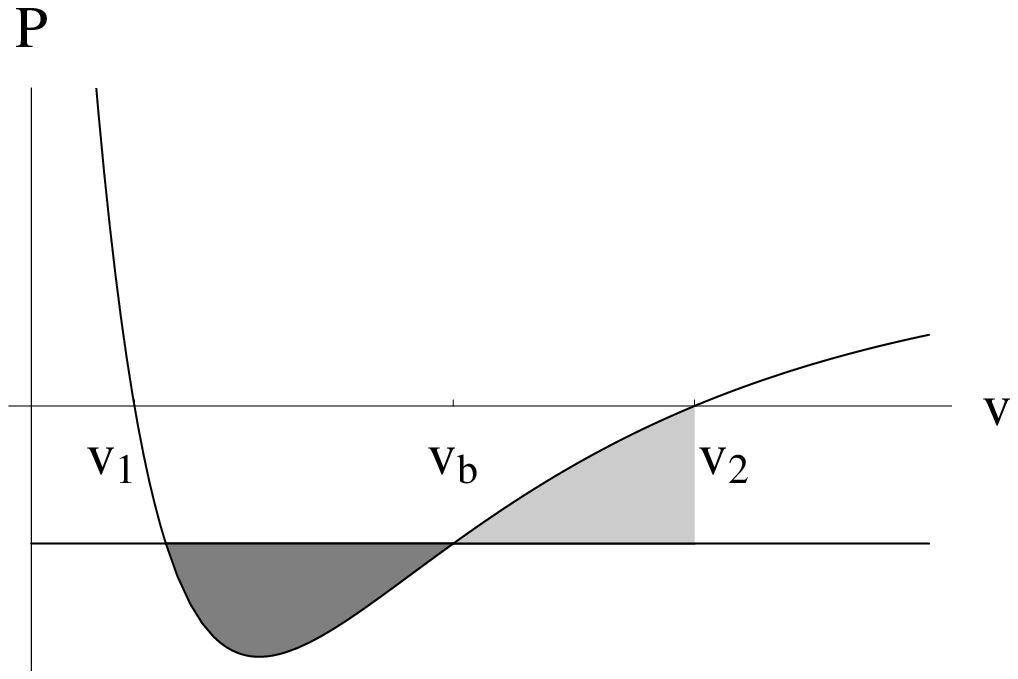}\\
  \caption{\footnotesize{A generic $P-v$ diagram showing an isotherm for $0<T<T_0=1/(6\sqrt{3}\pi |Q|)$ and an isobar $P=\text{const}<0$. The Maxwell's equal area law is replaced by the law expressing the equality of the dark grey and light grey areas, with $v_1$ and $v_2$ being the specific volumes where the isotherm intersects the $v$ axis and $v_b$ the largest specific volume where the isobar intersects the isotherm.}}\label{Fig2}
\end{figure}

For the Reissner-Nordstr\"om dS black hole, the pressure is still given by $P=P_{\La}\equiv -\La/8\pi<0$, since this is negative there is no way to observe criticality. By a similar reasoning to the one given in the previous paragraph, we conclude that $T_0$ is an upper-limit temperature for the charged dS black holes, however, in this case $Q$ and $M$ need not be huge.

For the Reissner-Nordstr\"om dS black hole where $P<0$~\eqref{nq9}, there is no oscillating part of the isotherm $0<T<T_0$ that is totally below the $v$ axis. In fact, the oscillating part starts below the $v$ axis and extends above it since, as $v$ increases, it is the term $T/v$ ($T>0$)~\eqref{eos} that becomes dominating. The transition from the small black hole to the large one may proceed along the isobar $P=\text{const}<0$, as shown in the generic Fig.~\ref{Fig2}, where the value of $P$ is selected on observing the law expressing the equality of the dark grey and light grey areas, with $v_1$ and $v_2$ being the specific volumes where the isotherm intersects the $v$ axis and $v_b$, the largest specific volume where the isobar intersects the isotherm. This replaces Maxwell's equal area law but mathematically is still expressed by the same line integral,
\begin{equation}\label{a2e}
    \oint v\,\dd P=0.
\end{equation}

For black holes where the pressure $P$~\eqref{nq9} may have equally both signs, the equal area law as defined in Fig.~\ref{Fig2}, and expressed by~\eqref{a2e}, applies to transitions with negative isobars.

In the case where $S=\pi s$, we have seen that the thermodynamic volume $V$ is equal to the geometric volume~\eqref{nq15}, as was the case treated in Ref.~\cite{Mann1}; that is, $V=\pi v^3/6$. Hence, the conclusion drawn in that reference, concerning the fact that the criticality cannot happen in the fixed $\Phi$ subensemble, holds in our general case too. In fact, in this case our EOS~\eqref{eos} takes the same form as Eq. (3.14) of Ref.~\cite{Mann1}\footnote{Equation~(3.14) of Ref.~\cite{Mann1} was misprinted.},
\begin{equation}\label{a3}
2\pi Pv^2-2\pi Tv+1-\Phi^2=0,
\end{equation}
which is derived from~\eqref{eos} on replacing $Q^2/v^2$ by $\Phi^2/4$. Equation~\eqref{a3} has no critical behavior. However, in the generalized case where $S\neq \pi s$, the relation between $V$ and $v$ is no longer of the form $V=\pi v^3/6$, as shown in Appendix B, and so the criticality may exist in the fixed $\Phi$ subensemble. For instance, if $S=\pi h(s,C)$, the EOS in the fixed $\Phi$ subensemble takes the form
\begin{equation}\label{a4}
2\pi Pv^2-2\pi Tv+1-\frac{4h(v^2/4,C)\Phi^2}{v^2}=0,
\end{equation}
which reduces to~\eqref{a3} if $h(s,C)=h(v^2/4,C)=s=v^2/4$.

\subsection{Special considerations: Critical exponents}

In the remaining part of this section, we restrict ourselves to black holes obeying the area law for entropy $S=\pi s$, which yields a geometric value for the thermodynamic volume $V=4\pi s^{3/2}/3$. We aim to derive general expressions for the critical exponents.

In terms of the thermodynamic variables used in this paper, the definitions of the critical exponents ($\al,\de,\bt,\ga$) along with their conditions of validity  are given by~\cite{book}
\begin{align}
&C_V\sim (\De T)^{-\al}[k_0+k_1\De T+\cdots] &(\De P=0),\nn\\
&\De P\sim (\De v)^{\de} &(\De T=0),\nn\\
\label{ce1}&v_2-v_1\sim (-\De T)^{\bt} &(\De P=0),\\
&\kappa_T\equiv -\frac{1}{V}\Big(\frac{\partial V}{\partial P}\Big)_{T} \sim (\De T)^{-\ga}&(\De P=0),\nn
\end{align}
where $\De P=P-P_c$, $\De v=v-v_c$, $\De T=T-T_c$, and ($k_0,k_1,\dots$) are constants.

The entropy $S$ being a function of $V$ only, any (partial) derivative of $S$ at constant $V$ is then 0. Particularly,
\begin{equation}\label{ce2}
C_V\equiv T\Big(\frac{\partial S}{\partial T}\Big)_{V}=0,
\end{equation}
this implies $\al =0$ and ($k_0=0,k_1=0,\dots$). Taking into account~\eqref{eosd} and~\eqref{a2}, we see that $\de =3$ since for $(v-v_c)/v_c\ll 1$, we obtain
\begin{equation}\label{ce3}
    \De P= -\frac{(\De v)^{3}}{3456\sqrt{6}\pi |Q|^5}+\cdots \quad (\De T=0).
\end{equation}

On the isotherm $T<T_c$ in the $P-v$ plane, where $-\De T/T_c\ll 1$, the point ($v_c,P_c$) is nearly a point of symmetry of the graph of the EOS~\eqref{eos} as shown in~\eqref{ce3}. This implies that the points $v_1$ and $v_2$ of Fig.~\ref{Fig1} are such that\footnote{By~\eqref{ce4}, Maxwell's equal area law is satisfied for $|\De v|/v_c\ll 1$ and $-\De T/T_c\ll 1$.}
\begin{equation}\label{ce4}
v_2-v_c\simeq v_c-v_1\qquad [(T-T_c)/T_c\ll 1].
\end{equation}
At the isobaric points ($v_1,P_1$) and ($v_2,P_2$), $P_2=P_1$, on the same isotherm $T<Tc$ (see Fig.~\ref{Fig1}), we obtain upon expanding~\eqref{eos} for $(v-v_c)/v_c\ll 1$ and $(T-T_c)/T_c\ll 1$,
\begin{multline}
\label{ce5}P_i-P_c=\frac{T-T_c}{v_c}-\frac{(v_i-v_c)^{3}}{6\times 24^2\sqrt{6}\pi |Q|^5}\\
-\frac{(T-T_c)(v_i-v_c)}{24Q^2}+\cdots \quad (i=1,2).
\end{multline}
Subtracting term by term these last two equations and using~\eqref{ce4} to reduce $(v_2-v_c)^{3}-(v_1-v_c)^{3}$ to
\begin{equation*}
(v_2-v_c)^{3}-(v_1-v_c)^{3}\simeq \frac{(v_2-v_1)^3}{4},
\end{equation*}
we arrive at
\begin{equation}
(v_2-v_1)^2\simeq 24^2\sqrt{6}\pi |Q|^3 (T_c-T),
\end{equation}
implying $\bt=1/2$. Using this last equation along with~\eqref{ce4}, we arrive at the other interesting result:
\begin{equation}
\frac{v_2-v_c}{v_c}\simeq \frac{v_c-v_1}{v_c}\simeq \sqrt{2}\sqrt{\frac{T_c-T}{T_c}}.
\end{equation}

With $V=\pi v^3/6$, write $\kappa_T$ as
\begin{equation*}
    \kappa_T=-\frac{3}{v}\,\frac{1}{(\partial P/\partial v)_T},
\end{equation*}
then evaluate $(\partial P/\partial v)_T$ from~\eqref{ce5}, after replacing ($P_i,v_i$) by ($P,v$) to get
\begin{equation*}
     \kappa_T \simeq 6\sqrt{6}|Q|(\De T)^{-1},
\end{equation*}
which results in $\ga =1$.

The four critical exponents are tabulated in Table~\ref{Tab1}. They satisfy the following thermodynamic scaling laws~\cite{AAMR}:
\begin{align}
&\alpha+2\beta+\gamma =2, &\alpha+\beta(\delta+1)=2, \nn\\
\label{ce6}&\gamma =\beta (\delta-1), &\gamma (\delta+1)=(2-\alpha)(\delta-1).
\end{align}

\begin{table}[!htb]
\centering
\caption{\footnotesize Critical exponents. \label{Tab1}} \vspace*{0.3cm}
\begin{tabular}{||l||c|c|c|c||} 
Symbol & $\al$ & $\de$ & $\bt$
& $\ga$  \\ \hline
Value & 0 & 3 & 1/2 & 1
\\ 
\end{tabular}
\end{table}

\section{Further applications \label{secfa}}

\subsection{Born-Infeld-AdS and Born-Infeld-dS solutions \label{secfa1}}

The Born-Infeld theory of general relativity~\cite{BI} admits a black hole solution in dS/AdS spacetime given by~\cite{BI1}-\cite{BI3}
\begin{equation}\label{bi1}
    f=1-\frac{2M}{r}-\frac{\La r^2}{3}+\frac{2b^2}{r}\int_r^{\infty}\Big[\sqrt{x^4+\frac{Q^2}{b^2}}-x^2\Big]\dd x.
\end{equation}
This metric has different types of black holes; the conditions of their existence, which depend on the values of ($M,Q,b,\La$), are discussed in Ref.~\cite{Mann2}. Here we assume that these conditions are met and that $f=0$ has a nonextremal event horizon.

We also assume that the cosmological constant $\La$ has both signs, which will allow us to thermodynamically treat equally the dS and the AdS cases. In the limit $b\to 0$, where $b$ is the strength of the electromagnetic field, the solution~\eqref{bi1} reduces to the Reissner-Nordstr\"om black hole in dS or AdS spacetime.

This solution can be brought to~\eqref{1.3} on adding and subtracting $Q^2/r^2$, so that $\sqrt{s}g(s)$ takes the form
\begin{align}
\sqrt{s}g&=\frac{\La s^{3/2}}{3}+\frac{Q^2}{\sqrt{s}}-2b^2\int_{\sqrt{s}}^{\infty}\Big[\sqrt{x^4+\frac{Q^2}{b^2}}-x^2\Big]\dd x\nn\\
\label{bi2}\quad &=\tfrac{Q^2}{\sqrt{s}}+\tfrac{(\Lambda -2 b^2) s^{3/2}}{3}+\tfrac{2b^2 s^{3/2}}{3}\, _2F_1\big(\tfrac{-3}{4},\tfrac{-1}{2};\tfrac{1}{4};\tfrac{-Q^2}{b^2
s^2}\big),
\end{align}
where $_2F_1$ is the hypergeometric function. The value of the pressure~\eqref{nq9} reads
\begin{equation}\label{bi3}
P(s)=\frac{b^2}{4 \pi }-\frac{\Lambda }{8 \pi }+\frac{Q^2}{8 \pi  s^2}-\frac{|b| \sqrt{Q^2+b^2 s^2}}{4 \pi  s}.
\end{equation}
The first two terms are constant contributions where each physical constant, ($\La ,b$), contributes a partial pressure. This is only apparent, and it does not mean that besides the constant pressure due the cosmological density, the nonlinear electromagnetic field contributes an additional constant pressure. In fact, in the limit $s\to\infty$, the term $b^2/(4 \pi)$ cancels out. The last two terms count for the effective contribution of the nonlinear electromagnetic field from which the linear part has been expelled.

For fixed ($Q,\La,b$), $P$ monotonically decreases as $s$ increases, for
\begin{equation*}
    P_{,s}=-\frac{Q^2}{4\pi s^3}\Big(1-\frac{|b|s}{\sqrt{Q^2+b^2 s^2}}\Big)<0.
\end{equation*}
With $\lim_{s\to0^+}P(s)=\infty$ and $\lim_{s\to\infty}P(s)=P_{\La}=-\La/(8\pi)$, we see that $P(s)$ is always positive for Born-Infeld-AdS black holes and that $P(s)$ vanishes one time then becomes negative for Born-Infeld-dS black holes. With $P_c>0$~\eqref{a2}, we conclude that
\begin{description}
  \item (1) criticality is attained for Born-Infeld-AdS black holes for some values of ($Q,\La,b$);
  \item (2) criticality is attained for small Born-Infeld-dS black holes where $P(s)$ remains positive;
  \item (3) there is no criticality for large Born-Infeld-dS black holes where $P(s)$ is negative.
\end{description}

For both Born-Infeld-AdS and Born-Infeld-dS black holes, there is an isobaric transition from the small to the large black hole where the value of the isobar $P=\text{const}>0$ is selected on applying Maxwell's equal area law. For Born-Infeld-dS black holes, a possible isobaric transition occurs at $P=\text{const}<0$ from a large black hole to a huge one, as shown in Fig.~\ref{Fig2}. When this is the case, the isobar $P=\text{const}<0$ is selected on applying the equal area law defined in Fig.~\ref{Fig2}.

These conclusions do not depend on the shape of the entropy function, for they are based on the EOS~\eqref{eos} which does not depend on the form of the function $S$.

Now if $S=\pi s$, then $V=4\pi s^{3/2}/3$ and all the above derived equations concerning the thermodynamic functions [Eqs.~\eqref{g2}, \eqref{g3}, \eqref{g3b}, \eqref{s2}, and~\eqref{s3}], critical values and law of corresponding states [Eqs.~\eqref{a2}, \eqref{a2a}, and~\eqref{a2b}], critical exponents (Table~\ref{Tab1}) and related conclusions and figures remain valid.

\subsection{Martinez-Troncoso-Zanelli solutions}

Martinez-Troncoso-Zanelli (MTZ) black holes are spherically symmetric solutions to the Einstein equations with a conformally coupled scalar field~\cite{MTZ}. If $V(\phi)=-\bt^2\phi^4$ is the scalar field self-interaction potential, where $\bt^2$ is a coupling constant,
\begin{equation}\label{MTZ1}
\bt^2\geq 2\pi\La/9,
\end{equation}
the metric function $f$ is given by
\begin{equation}\label{MTZ2}
f=\Big(1-\frac{M}{r}\Big)^2-\frac{\La r^2}{3},
\end{equation}
and the scalar field by
\begin{equation*}
    \phi(r)=\sqrt{\frac{\La}{6\bt^2}}\,\frac{M}{M-r},
\end{equation*}
which is regular on the horizons. The latter are nonextremal provided
\begin{equation}\label{MTZ2b}
\La >0\quad \text{ and }\quad M<\frac{1}{4}\sqrt{\frac{3}{\La}}.
\end{equation}

The saturation in~\eqref{MTZ1} corresponds to the neutral solution, and the condition $\bt^2> \La/36$ implies the existence of charged solutions where the charge is related to the mass by
\begin{equation}\label{MTZ3}
Q^2=M^2-\frac{2\pi\La}{9\bt^2}\,M^2.
\end{equation}

The metric~\eqref{MTZ2} can be brought to~\eqref{1.3} on adding and subtracting $Q^2/r^2$, so that $\sqrt{s}g(s)$ takes the form
\begin{equation}
\sqrt{s}g=\frac{\La s^{3/2}}{3}-\frac{2\pi\La M^2}{9\bt^2\sqrt{s}},
\end{equation}
where we have used~\eqref{MTZ3}. The value of the pressure~\eqref{nq9} reads
\begin{equation}\label{MTZ4}
P(s)=-\frac{\Lambda }{8 \pi }-\frac{\Lambda M^2}{36 s^2 \beta ^2}.
\end{equation}
Using~\eqref{MTZ2b}, this is always increasing and negative, and $\lim_{s\to\infty}P(s)=P_{\La}=-\La/(8\pi)$. Thus, criticality is not attained. A possible isobaric transition occurs at $P=\text{const}<0$ from a large black hole to a larger one [as shown in Fig.~\ref{Fig2}], provided~\eqref{MTZ2b} remains satisfied. When this is the case, the isobar $P=\text{const}<0$ is selected on applying the equal area law defined in Fig.~\ref{Fig2}.

These conclusions do not depend on the shape of the entropy function, for they are based on the EOS~\eqref{eos} which does not depend on the form of the function $S$.

The entropy of the MTZ black holes is not given by the area law $S=\pi s$~\cite{MTZ2}; thus, some of the above determined quantities, as the critical exponents (Table~\ref{Tab1}), are not valid.

\subsection{Regular black holes \label{secfa4}}

\subsubsection{Ay\'on-Beato--Garc\'{\i}a black hole}

The metric function $f$ of the Ay\'on-Beato--Garc\'{\i}a regular black hole~\cite{reg} is given by
\begin{equation}\label{rb1}
    f=1-\frac{2Mr^2}{(r^2+Q^2)^{3/2}}+\frac{Q^2r^2}{(r^2+Q^2)^{2}},
\end{equation}
where $M$ and $Q$ are the mass and the electric charge. Bringing this metric function $f$ to the form~\eqref{1.3} we obtain
\begin{equation}\label{rb2}
\sqrt{s}g=-2M+\frac{Q^2}{\sqrt{s}}+\frac{2Ms^{3/2}}{(s+Q^2)^{3/2}}-\frac{Q^2s^{3/2}}{(s+Q^2)^{2}},
\end{equation}
which depends on $M$ and $Q^2$. The pressure formula~\eqref{nq9} yields
\begin{equation}\label{rb3}
\hspace{-1mm}P(s)=\frac{Q^4 (6 s^2+3 Q^2 s+Q^4)-6 M Q^2 s^2 \sqrt{s+Q^2}}{8 \pi  s^2 (s+Q^2)^{3}}.
\end{equation}

Knowing that the extremality condition for this black hole is $M^2/Q^2\simeq 2.49$~\cite{AzregR}, we consider only the black hole configuration corresponding to $M^2/Q^2> 2.49$. With $\lim_{s\to 0^+}P(s)=\infty$ and $\lim_{s\to \infty}P(s)=0^-$, we see that $P(s)$ may assume any positive value and that it has a negative absolute minimum value. First, criticality is attained for $M^2/Q^2> 2.49$. Second, contrary to the Reissner-Nordstr\"om black hole, where the isobaric transition from small to large black holes occurs at $P=0$, this transition occurs at positive pressures as well as at negative ones, as shown in Fig.~\ref{Fig2}, provided the value of the isobaric pressure remains larger than the negative absolute minimum value.

To our knowledge, there are no corrections to the area law for entropy, so we can assume $S=\pi s$, and thus all the above formulas derived under this hypothesis remain valid.

\subsubsection{Charged regular cores}

In Ref.~\cite{IrinaD} the properties of a regular core with the dS interior and variable mass parameter were investigated. Its metric function $f$ reads
\begin{equation}\label{rc1}
f=1-\frac{2 M}{r} \Big(1- \e^{-\frac{r^3}{2 r_0{}^2 M}}\Big).
\end{equation}
In the limit $r\to 0$, $f$ behaves as a dS solution: $f= 1-r^2/r_0{}^2+\mathcal{O}(r^5)$ with $r_0{}^2=3/\La$.

Since the electric charge is 0, the corresponding EOS~\eqref{eos} does not show any critical behavior (see Sec.~\ref{secE} for a similar discussion). For this case $Q=0$, transitions from small to large black holes occur for positive pressures only since the equivalent of Fig.~\ref{Fig1} would be a plot extending to infinitely negative pressures, reaching a positive absolute maximum value, then decreasing and finally approaching 0 from above for large values of $v$. So, coexistence exists only for positive pressures.

We rather like to investigate the charged counterpart of~\eqref{rc1} which we assume has the form
\begin{align}
\label{rc2}f&=1-\frac{2 M}{r} \Big(1-\e^{-\frac{\Lambda  r^3}{3 M}}\Big)+\frac{Q^2}{r^2} \Big(1-\e^{-\frac{\Lambda  r^4}{3 Q^2}}\Big)\\
\label{rc3}&=1-\frac{2 M}{r}+\frac{Q^2}{r^2}+\frac{2 M}{r}\,\e^{-\frac{\Lambda  r^3}{3 M}}-\frac{Q^2}{r^2} \,\e^{-\frac{\Lambda  r^4}{3 Q^2}}.
\end{align}
In the limit $r\to 0$, $f$ behaves as a dS solution,
\begin{equation*}
f= 1-\La r^2/3+\mathcal{O}(r^5),
\end{equation*}
and in the limit $r\to\infty$, $f$ approaches a Reissner-Nordstr\"om solution,
\begin{align*}
f\simeq 1-\frac{2 M}{r}+\frac{Q^2}{r^2}.
\end{align*}

In the context of our approach, we may consider the solution either as a variable mass regular core~\eqref{rc2} or as a constant mass one~\eqref{rc3}; in both treatments, we reach the same conclusions. In the latter case, we obtain from~\eqref{rc3}
\begin{equation*}
\sqrt{s}g=\frac{Q^2}{\sqrt{s}}\,\e^{-\frac{\Lambda  s^2}{3 Q^2}}-2M\e^{-\frac{\Lambda  s^{3/2}}{3 M}},
\end{equation*}
and from~\eqref{nq9} [with $M(s)_{,s}=0$)]
\begin{equation}\label{rc4}
P(s)=\frac{Q^2\e^{-\frac{\La s^2}{3 Q^2}}}{8 \pi  s^2}+\frac{2 \Lambda  \e^{-\frac{\La s^2}{3 Q^2}}-3 \Lambda  \e^{-\frac{\La s^{3/2}}{3M}}}{12 \pi }.
\end{equation}

Extremality cannot be determined analytically; however, numerically, we confirm the existence of two separate roots of $f(r)=0$ for some set of the parameter ($M,Q,\La$).

This case is very similar to the case treated in the previous section, for $\lim_{s\to 0^+}P(s)=\infty$ and $\lim_{s\to \infty}P(s)=0^-$. This implies that $P(s)$ may assume any positive value and that it has a negative absolute minimum value. Thus, criticality is attained when the black hole solution exists. Similarly to the previous case, the isobaric transition from small to large black holes occurs at positive pressures as well as at negative ones provided the value of the pressure remains larger than the negative absolute minimum value.

We assume $S=\pi s$, as is the case with the neutral solution~\eqref{rc1}, and thus all the above formulas derived under this hypothesis remain valid.

This thermodynamic behavior, of the two charged regular black holes considered here, extends most likely to all known charged regular black holes.

\section{Thermodynamic potentials in the presence of quintessence \label{secq}}

Quintessence could also be introduced via an effective energy-stress tensor, $T_{\text{eff}\,\mu}{}^{\nu}\equiv T_{q\;\mu}{}^{\nu}$. If the mass is constant and quintessence is present alone with no other external forces,
\begin{equation*}
    f(r)=1-\frac{2M}{r}+\frac{Q^2}{r^2}-g_q(r),
\end{equation*}
the only modification one needs to introduce is a change in the rhs of~\eqref{nq8} so that the partial pressure $P_q$ due to quintessence is given by\footnote{It is usually admitted that $-1<\om<0$, but the teleparallel dark energy cosmological model~\cite{para1,para2} allows for $\om<-1$~\cite{para3}.}
\begin{equation}\label{q8}
\hspace{-3mm}P_q=\om \frac{\ep_q}{8\pi}\Big|_{u=s}=\om\frac{[\sqrt{s}g_q(s,C_q)]_{,s}}{4\pi \sqrt{s}}\quad (-1<\om<0).
\end{equation}
Note that $-P_q$, which is proportional to $T_{t}{}^{t}$, is the average value over the angles of the diagonal spatial components of the SET~\cite{AzregE,Kis,AAR}. The corresponding thermodynamic volume $V_q$ still satisfies~\eqref{nq13} where we drop the term $2M$ from the rhs,
\begin{equation*}
\frac{V_{q,\,s}P_q}{\pi}=
-\frac{(\sqrt{s}g)_{,s}}{2\pi}.
\end{equation*}
This yields, using~\eqref{q8}, $V_{q,\,s}=-2\pi \sqrt{s}/\om$, so that
\begin{equation}\label{q15}
    V_q(s)=-\frac{4\pi}{3\om}s^{3/2}.
\end{equation}
The expression of $H$ and the corresponding formulas~\eqref{nq16} become
\begin{align}
&H(S,Q,P)=\frac{\sqrt{s}}{2}+\frac{Q^2}{2\sqrt{s}}-\frac{4\pi}{3\om}s^{3/2}P_q,\nn\\
\label{q16}&V_q\equiv \Big(\frac{\partial H}{\partial P_q}\Big)_{S,Q}=-\frac{4\pi}{3\om}s^{3/2},\\
&T=\frac{1}{4\pi \sqrt{s}}-\frac{Q^2}{4\pi s^{3/2}}-2\sqrt{s}\frac{P_q}{\om},\nn\\
&\dd H = T \dd S+A\dd Q+V_q\dd P_q.\nn
\end{align}

Generalization to the case where, besides quintessence, other external forces are also present is straightforward due to the linearity of~\eqref{nq13}. In this case the metric function $f$, including a variable-mass term, takes the form
\begin{equation}\label{ggq}
     f(r)=1-\frac{2M(r)}{r}+\frac{Q^2}{r^2}-\sum_ig(r,C_i)-g_q(r,C_q).
\end{equation}
The general expressions extending~\eqref{g2} to include quintessence read
\begin{align}
&H(S,Q,P,P_q)=\frac{\sqrt{s}}{2}+\frac{Q^2}{2\sqrt{s}}+\frac{4\pi}{3}s^{3/2}P-\frac{4\pi}{3\om}s^{3/2}P_q,\nn\\
&P=\sum_i P_i-\frac{M(s)_{,s}}{2\pi\sqrt{s}},\quad P_i=-\frac{[\sqrt{s}g_i(s,C_i)]_{,s}}{4\pi \sqrt{s}},\nn\\
\label{q17}&V\equiv \Big(\frac{\partial H}{\partial P}\Big)_{S,Q,P_q}=\frac{4\pi}{3}s^{3/2},\\
&P_q=\om\frac{[\sqrt{s}g_q(s,C_q)]_{,s}}{4\pi \sqrt{s}},\;\, V_q\equiv \Big(\frac{\partial H}{\partial P_q}\Big)_{S,Q,P}=-\frac{4\pi}{3\om}s^{3/2},\nn\\
&T=\frac{1}{4\pi \sqrt{s}}-\frac{Q^2}{4\pi s^{3/2}}+2\sqrt{s}\Big(P-\frac{P_q}{\om}\Big),\nn\\
&\dd H = T \dd S+A\dd Q+V\dd P+V_q\dd P_q.\nn
\end{align}
This set of equations generalizes the results concerning the Reissner-Nordstr\"om--de Sitter black hole surrounded by quintessence~\cite{AzregE} to all spherically symmetric black holes, whether they are solution to $G_{\mu}{}^{\nu}=8\pi T_{\mu}{}^{\nu}$ or not.

The EOS forms~\eqref{g5} and~\eqref{eos} generalize to [the fifth line in~\eqref{q17}]
\begin{align}
\label{q18}P-\frac{P_q}{\om}&=\frac{T}{2\sqrt{s}}-\frac{1}{8\pi s}+\frac{Q^2}{8\pi s^2}\\
&=\frac{T}{v}-\frac{1}{2\pi v^2}+\frac{2Q^2}{\pi v^4}.\nn
\end{align}

To generalize the other thermodynamic potentials, we need first generalize the Smarr formula~\eqref{g3}. Since $H$~\eqref{q17} is homogeneous in ($S,Q^2,P^{-1},P_q^{-1}$) of order $1/2$, this becomes
\begin{equation}
  H=2T S+AQ-2VP-2V_qP_q,
\end{equation}
which, along with $U=H-PV-V_qP_q$, $F=U-TS$, and $G=H-TS$, yields
\begin{align}
F&=T S+AQ-3VP-3V_qP_q\nn\\
\label{qs3}&=T S+AQ-3V\Big(P-\frac{P_q}{\om}\Big),\\
G&=T S+AQ-2VP-2V_qP_q\nn\\
\label{qs4}&=T S+AQ-2V\Big(P-\frac{P_q}{\om}\Big),
\end{align}
where we have used $V_q=-V/\om$. We see that the expression~\eqref{g3b} of $U$ remains unchanged. The expression~\eqref{s2} also remains unchanged as this is clear from~\eqref{q18} and~\eqref{qs3}. This confirms the statement made in Ref.~\cite{AzregE}: ``The internal energy of any static charged black hole, with possibly a variable mass parameter, does depend explicitly only on the intrinsic properties of the black hole." The final expression of $G$ reads
\begin{equation}\label{qs5}
G(T,P,P_q,Q)=\frac{\sqrt{s}}{4}-\frac{2\pi s^{3/2}}{3}\Big(P-\frac{P_q}{\om}\Big)+\frac{3Q^2}{4\sqrt{s}},
\end{equation}
where $s$ is an implicit solution to the EOS~\eqref{q18} expressed in terms of ($T,P-P_q/\om,Q$).\\

\paragraph*{\textbf{Application.}}

In four-dimensional spacetime, a spherical symmetric Reissner-Nordstr\"om black hole plunged into the field of a spherical symmetric quintessence has its metric function $f$ given by~\cite{Kis}
\begin{equation*}
f(r)=1-\frac{2M}{r}+\frac{Q^2}{r^2}-\frac{2c}{r^{3\om+1}}, -1< \om<0 \text{ and } c>0.
\end{equation*}
Here we are using the notation of~\cite{AzregE,AAR}. If we include a cosmological constant, the metric takes the general form
\begin{multline}\label{f1}
f(r)=1-\frac{2M}{r}+\frac{Q^2}{r^2}-\frac{\La r^2}{3}-\frac{2c}{r^{3\om+1}},\\\text{ with } -1< \om<0, \text{ and } c>0.
\end{multline}
This is of the form~\eqref{ggq} with $g=g_{\La}=\La r^2/3$ and $g_q=2c/r^{3\om+1}$ yielding
\begin{equation}\label{f2}
    P=P_{\La}=-\frac{\La}{8\pi}\quad\text{ and }\quad P_q=-\frac{3\om^2c}{4\pi s^{3(\om+1)/2}}.
\end{equation}

If $\La >0$ or $\La =0$ (no cosmological constant), the lhs of~\eqref{q18},
\begin{equation*}
    -\frac{\La}{8\pi}+\frac{3\om c}{4\pi s^{3(\om+1)/2}},
\end{equation*}
is negative, so criticality is not observed, but transitions from small to large black holes are possible as shown in Fig.~\ref{Fig2}. If $\La <0$, the lhs of~\eqref{q18} is positive only for large $v(=2\sqrt{s})$. So criticality may be attained for large values of $v$ provided~\eqref{a2bb} and~\eqref{a2c} hold; that is, provided $M$ is large enough. But this is likely the case, since large $v$ corresponds to large $M$. These conclusions remain valid for all values of $\om$.

The special case $\La <0$ and $\om=-2/3$ was treated analytically in Ref.~\cite{Li} in the $P_{\La}=\text{const}>0$ ensemble without discussing further constraints on criticality. The conclusion remains qualitatively the same as the one made here since, in this ensemble, criticality is also subject to the constraints~\eqref{a2bb} and~\eqref{a2c}. In this ensemble, the enthalpy is equal to the mass parameter $M$~\cite{Li} while the enthalpy in our $P_{\La}-P_q/\om=\text{const}$ ensemble is given by the first line in~\eqref{q17} with $P=P_{\La}$. So the two enthalpy functions are related by~\cite{AzregE}
\begin{equation}
M=H+\frac{4\pi}{3}\frac{\om+1}{\om^2}s^{3/2}P_q.
\end{equation}

Two other illustrative examples of such a relation between the enthalpy functions of two different ensembles or systems are provided in the following section.

\section{The $\pmb P$-$\pmb V$ term in the early literature on black hole thermodynamics \label{secE}}

The thermodynamics of charged black holes is usually treated in such a way to highlight the term $A\dd Q$ (which is the work done by electric forces) in comparison with classical thermodynamics. Within the framework of our approach, we show that this term could be absorbed in a $P$-$V$ term; these approaches correspond to different ensembles.

For the sake of example, we first consider the Reissner-Nordstr\"om black hole. So far we have treated the electromagnetic field as an intrinsic property of the black hole. Now, we treat the field as an external force exerting a pressure on the horizon; that is, we consider the Reissner-Nordstr\"om black hole as a neutral ($Q=0$) Schwarzschild hole in an external electric field. In this case, \eqref{1.3} takes the form
\begin{equation}\label{e1}
    f(r)=1-\frac{2M}{r}-g_e(r,Q),\quad g_e(r,Q)=-\frac{Q^2}{r^2}.
\end{equation}
The pressure $P_e$ due to the external electric field is derived from~\eqref{nq9},
\begin{equation}\label{e2}
    P_e=-\frac{Q^2}{8\pi s^2}<0.
\end{equation}
Using~\eqref{nq16} (with $Q=0$), we obtain the enthalpy and temperature of a Schwarzschild black hole immersed in a radial electric field
\begin{align}
&H_e(S,P_e)=\frac{\sqrt{s}}{2}+\frac{4\pi}{3}s^{3/2}P_e,\nn\\
\label{e3}&V\equiv \Big(\frac{\partial H_e}{\partial P_e}\Big)_{S}=\frac{4\pi}{3}s^{3/2},\\
&T\equiv \Big(\frac{\partial H_e}{\partial S}\Big)_{P_e}=\frac{1}{4\pi \sqrt{s}}+2\sqrt{s}P_e,\nn\\
&\dd H_e = T \dd S+V\dd P_e.\nn
\end{align}

Using~\eqref{e2}, it is easy to check that the temperature $T$~\eqref{e3} is that of a Reissner-Nordstr\"om black hole. Using this in $\dd (H_e-M)=V\dd P_e-A\dd Q\neq 0$\footnote{Here we have used the fact that the differential of the mass parameter is related to the differentials of $S$ and $Q$ by $\dd M=T\dd S+A\dd Q$. In our context, this is seen as a pure mathematical relation not related to ensembles.}, we see that the enthalpy is different from the mass parameter. In fact, we directly evaluate $H_e$ on substituting~\eqref{e2} in the first line of~\eqref{e3},
\begin{equation}\label{e4}
 H_e=\frac{\sqrt{s}}{2}-\frac{Q^2}{6\sqrt{s}}.
\end{equation}
This is to say that the enthalpy of the Reissner-Nordstr\"om black hole [derived from~\eqref{nq16} setting $P=0$ but $Q\neq 0$], which satisfies $H=M$, is not the enthalpy of the Schwarzschild black hole immersed in an electric field having the same value and temperature as that of the Reissner-Nordstr\"om black hole. This seems obvious, noticing that the enthalpy is used in processes where the pressure is held constant, and that the thermodynamic processes $P=\text{const }(=0)$ and $P_e=\text{const }(\neq 0)$ are very different. In the former process, the parameters are subject to no constraint, and in the latter one $Q$ and $S$ are subject to $Q^2=\text{const}\,s^2$~\eqref{e2}.

The general expression providing the enthalpy change between two systems, where the linear part of the electromagnetic field is seen as internal or external to the system, is
\begin{equation}\label{ch}
    \frac{\sqrt{\pi}Q^2}{2\sqrt{S}}-VP_e\qquad (V=4\pi s^{3/2}/3).
\end{equation}
This is the potential electric energy of the internal linear part of the electromagnetic field minus the pressure work of the external linear part.

In the $P=0$ ensemble, while the EOS~\eqref{eos} is still satisfied, no criticality is observed for $P=0<P_c$. The EOS of the $P=P_e$ ensemble reduces to
\begin{equation}\label{e5}
P_e=\frac{T}{v}-\frac{1}{2\pi v^2},
\end{equation}
and this has no critical points; the graph of $P_e$ versus $v$ is a plot extending to infinitely negative pressures, reaching a positive absolute maximum value, then decreasing and finally approaching 0 from above for large values of $v$. In the $P=0$ ensemble, there is an isobaric transition from a small black hole to a large one at $P=0$ and at some temperature $T<T_0<T_c$ selected to satisfy Maxwell's equal area law. There is no such transition in the $P=P_e$ ensemble since the rhs of~\eqref{e5} is monotonically increasing when its value is negative; that is, there is no coexistence for negative pressures.

Another example is the black hole of the Born-Infeld theory, which was treated in Sec.~\ref{secfa1}, considering the linear part of the electromagnetic field as an internal property and the nonlinear part as an external force. It could be treated considering all the electromagnetic field, including its linear part, as an external force. In that case, the terms $Q^2/(8 \pi  s^2)$ and $2Q^2/(\pi  v^4)$ in the rhs of~\eqref{bi3} and~\eqref{eos}, respectively, would disappear. The remaining EOS, similar to~\eqref{e5}, would have no criticality, and there would be no isobaric transitions from small to large black holes. The enthalpy change is given by~\eqref{ch}.

We see that considering the linear part of the electromagnetic field as (internal) part of the thermodynamic system and treating the remaining part of the field as an external force enriches the thermodynamic investigation of black hole solutions.

\section{Conclusion \label{secc}}

The thermodynamics of black holes may be described by different choices of variables, which means that the notion of ensembles for black holes is larger than that in classical thermodynamics. These choices range from parametric to physical, in that some descriptions make use of parameters leading to extra terms, such as $\Ta \dd C$, not having a direct physical meaning, and some other descriptions rely on physical variables leading to thermodynamic entities with which one is familiar, such as $P$-$V$ terms.

Based on the idea that the linear contribution of the electromagnetic field is seen as an intrinsic property of the thermodynamic system, we have made a judicious choice of the thermodynamic variables by redefining theme, particularly a judicious choice of the pressure, by which we derived expressions for the thermodynamic potentials, thermodynamic volume, and critical exponents that apply to all black holes obeying the area law for entropy.

Whether the area law for entropy applies or not, another fact derived in this work is that criticality may exist only for charged black holes (see last paragraph of this section). Using the universal EOS, we have shown that the transitions from small to large black holes occur for positive as well as negative pressures depending on the nature of the, neutral or charged, black hole solution, contrary to classical thermodynamics where these transitions take place for positive pressures only.

For many black hole thermodynamic applications, the mass parameter does not seem to be the appropriate thermodynamic potential.

A common feature to charged regular black holes is that they show critical behavior, contrary to Reissner-Nordstr\"om singular ones, and their isobaric transitions from small to large black holes occur for both signs of the pressure. The inclusion of the $P$-$V$ term has removed any inconsistency in their thermodynamic treatment since this is better described using the enthalpy (which is different from the mass parameter) and applying the area law for entropy.

Finally, the inclusion of quintessence has shown that the thermodynamic volume for nonrotating solution may be different from the geometric volume. Even if no quintessence force is present, this conclusion extends to the case where the area law for the entropy no longer holds. This discrepancy between the two volumes is already known for rotating solutions~\cite{Dolan}.

We emphasize that the approach introduced here applies to any solution to $G_{\mu}{}^{\nu}=8\pi T_{\mu}{}^{\nu}$, or more generally, to $G_{\mu}{}^{\nu}=8\pi T_{\text{eff}\,\mu}{}^{\nu}$ provided the metric components are such that $g_{tt}=-1/g_{rr}$ and $g_{\ta\ta}=g_{\varphi\varphi}/\sin^2\ta=-r^2$. A generalization to metrics of the form $g_{tt}=-1/g_{rr}$ and $g_{\ta\ta}=g_{\varphi\varphi}/\sin^2\ta\neq -r^2$, to include other type of black holes, as phantom ones~\cite{gerard,thermo,AAMR} and those derived in scalar tensor theories, may be achieved in a subsequent work.

Recently, the authors of Ref.~\cite{van1} have shown, via the inclusion of a $P$-$V$ term, that not only neutral black holes may have critical phenomena, but that some special near horizon nonvacuum solutions to $G_{\mu\nu}+\La g_{\mu\nu}=8\pi T_{\mu\nu}$, with $\La <0$ and $P=-\La /(8\pi)$, have their EOS, which describes the thermodynamics of the event horizon, identical to the EOS of a Van der Waals fluid. The investigation was later extended to any spacetime dimension $d$ and to different horizon topologies (spherical, planar, and hyperbolic)~\cite{van2}. It was shown that the SET sourcing the field equations is an anisotropic fluid and that all three (weak, strong, and dominant) energy conditions are satisfied in a neighborhood of the horizon.

\section*{Appendix A: Relationship between enthalpy and mass parameter \label{secapdxa}}
\renewcommand{\theequation}{A.\arabic{equation}}
\setcounter{equation}{0}

With $f(r)$ given by~\eqref{1.3}, we have seen that $f(s)=0$ implies $M(s)=E(s)$ [but $M(s)_{,s}\neq E(s)_{,s}$]
\begin{equation}\label{B1}
    M(s)=E(s)=\frac{\sqrt{s}}{2}+\frac{Q^2}{2\sqrt{s}}-\frac{\sqrt{s}g(s)}{2}.
\end{equation}
We want to compare the value of $M(u)$ at $s$, that is, $E(s)$ as given in~\eqref{B1} with the expression of $H$ which is the first line in~\eqref{nq16} with $P$ given by~\eqref{nq9}. When $E(s)=H$ is true, we write $H=M$, but this does not mean $H(u)=M(u)$. This comparison leads to
\begin{equation}\label{B2}
2\sqrt{s}M_{,s}=g-sg_{,s}.
\end{equation}
Since here $M_{,s}$ and $g_{,s}$ have been obtained upon first deriving the expressions of $M(u)$ and $g(u)$ with respect to $u$ [see Eqs.~\eqref{nq6} and~\eqref{nq8}], then replacing $u$ by $s$, so we can rewrite~\eqref{B2} using the coordinate $u$,
\begin{equation}\label{B3}
2\sqrt{u}M_{,u}=g-ug_{,u},
\end{equation}
yielding
\begin{equation}\label{B4}
    M(r)=K_1+\int^{r^2}\frac{g(u')-u'g(u')_{,u'}}{2\sqrt{u'}}\,\dd u',
\end{equation}
where $K_1$ is a constant of integration.

If $M$ is constant, Eq.~\eqref{B3} results in
\begin{equation}\label{B4b}
    g=K_2 u=K_2 r^2,
\end{equation}
where $K_2$ is a constant. The case $K_2=0$ corresponds to the Reissner-Nordstr\"om black hole. If $K_2\neq 0$, we identify $K_2u$ with the de Sitter or anti--de Sitter term and $K_2$ with $\La/3$ ($\La >0$ or $\La <0$, respectively). Conversely, if $g\propto u$, then~\eqref{B3} yields $M=\text{const}$. We have thus shown that, for a Reissner-Nordstr\"om black hole, a charged dS, and a charged AdS black hole with a constant mass parameter, $H=M$.

For a Reissner-Nordstr\"om black hole, a charged dS or AdS black hole with a variable mass parameter where $g=K_2 r^2$, $H\neq M$ [$H\neq E(s)$]; rather,
\begin{equation}\label{B5}
    H=E(s)-\frac{2sM(s)_{,s}}{3},
\end{equation}
where $M(s)_{,s}=M(u)_{,u}|_{u=s}$. This shows that for the dS solutions used as regular cores~\cite{irina}-\cite{conf}, where the mass parameter is not constant, the latter is different from the enthalpy and related to it by~\eqref{B5}, where $E(s)$ is the value of the mass parameter on the horizon.

We have just seen that if the mass parameter is constant, the charged dS and AdS black holes (including Reissner-Nordstr\"om black holes) satisfy~\eqref{B4}; other black hole solutions may satisfy~\eqref{B4} provided the mass parameter depends on $r$. When this is the case, we have $H=M$ and the pressure~\eqref{nq9} reduces to
\begin{equation}\label{B6}
P=-\frac{3g(s)}{8\pi s},
\end{equation}
generalizing the expression of the pressure for the charged dS and AdS black holes, which is $P_{\La}=-\La/(8\pi)$ with $g_{\La}(s)=\La s/3$~\eqref{B4b}.

\section*{Appendix B: Thermodynamic volume for the case $\pmb{S=\pi h(s,C)}$\label{secapdxb}}
\renewcommand{\theequation}{B.\arabic{equation}}
\setcounter{equation}{0}

Consider Eq.~\eqref{g4b} where we assume that the thermodynamic volume $V$ is a function of ($S,C$). Differentiating~\eqref{g4b} with respect to $S$ then replacing $S$ by $\pi h(s,C)$, we obtain
\begin{equation}\label{A1}
\Big(\frac{\partial H}{\partial S}\Big)_{Q,P}=W+V_{,S}P,
\end{equation}
where we have set
\begin{equation*}
W\equiv \frac{1}{4\pi \sqrt{h}}-\frac{Q^2}{4\pi h^{3/2}}.
\end{equation*}
Similarly, setting
\begin{equation*}
w\equiv \frac{1}{4\pi \sqrt{s}}-\frac{Q^2}{4\pi s^{3/2}},
\end{equation*}
the requirement~\eqref{nq10a} yields
\begin{equation}\label{A2}
V_{,S}=2\sqrt{s}+\frac{w-W}{P},
\end{equation}
where we have used the expression of $T$ as given in~\eqref{g2}.

To determine $V_{,s}$ in terms of $V_{,S}$, we use~\eqref{nq9} where $P$ is kept constant; if $g$ depends on $Q$, then this is also kept constant. We are led to, upon differentiating $4\pi P \sqrt{s}=-[\sqrt{s}g(s,C)+2M(s)]_{,s}$,
\begin{equation}\label{A3}
C_{,s}\equiv \Big(\frac{\partial C}{\partial s}\Big)_{P,Q}=\frac{2\pi P-\sqrt{s}(\sqrt{s}g+2M)_{,ss}}{\sqrt{s}(\sqrt{s}g+2M)_{,sC}},
\end{equation}
where $(\sqrt{s}g+2M)_{,sC}$ can be written as $(\sqrt{s}g)_{,sC}$ if $M$ does not depend on $C$. Using this expression of $C_{,s}$ in
\begin{equation}\label{A4}
\dd S=\pi (h_{,s}+C_{,s}h_{,C})\dd s,
\end{equation}
we arrive at
\begin{equation}\label{A5}
V_{,s}=2\pi\sqrt{s}(h_{,s}+C_{,s}h_{,C})+\frac{\pi (h_{,s}+C_{,s}h_{,C})(w-W)}{P},
\end{equation}
where, in Eqs.~\eqref{A3} and~\eqref{A5}, $P(s)$ is given by the rhs of~\eqref{nq9}. Notice that the rhs of~\eqref{A5} reduces to~\eqref{nq14} if $h(s,C)\equiv s$, which yields $h_{,C}\equiv 0$ and $w=W$. The differential Eq.~\eqref{A5} is not generally tractable analytically.



\begin{thebibliography}{99}

\bb{ex0}M.M.~Caldarelli, G.~Cognola, and D.~Klemm,
\emph{Thermodynamics of Kerr-Newman-AdS black holes and conformal field theories},
Class. Quantum Grav. \textbf{17} 399 (2000). arXiv:hep-th/9908022.

\bb{Wang-Wu}S.~Wang, S.Q.~Wu, F.~Xie, and L.~Dan,
\emph{The first laws of thermodynamics of the (2+1)-dimensional BTZ black holes and Kerr-de Sitter spacetimes},
Chin. Phys. Lett. \textbf{23} 1096 (2006). arXiv:hep-th/0601147.

\bb{Sekiwa}Y.~Sekiwa,
\emph{Thermodynamics of de Sitter black holes: Thermal cosmological constant},
Phys. Rev. D \textbf{73} 084009 (2006). arXiv:hep-th/0602269.

\bb{Wang2}S.~Wang,
\emph{Thermodynamics of Schwarzschild de Sitter spacetimes: Variable cosmological constant},
arXiv:gr-qc/0606109.

\bb{nuc}G.L.~Cardoso and V.~Grass,
\emph{On five-dimensional non-extremal charged black holes and FRW cosmology},
Nucl. Phys. B \textbf{803} 209 (2008). arXiv:0803.2819 [hep-th].

\bb{ex1}D.~Kastor, S.~Ray, and J.~Traschen,
\emph{Enthalpy and the Mechanics of AdS Black Holes},
Class. Quantum Grav. \textbf{26} 195011 (2009). arXiv:0904.2765 [hep-th].

\bb{Dolan2}B.P.~Dolan,
\emph{Pressure and volume in the first law of black hole thermodynamics},
Class. Quantum Grav. \textbf{28} 235017 (2011). arXiv:1106.6260 [gr-qc].

\bibitem{Dolan}B.P.~Dolan, ``Where is the PdV in the First Law of Black Hole Thermodynamics?" in \textit{Open Questions in Cosmology}, edited by G.J.~Olmo (\href{http://www.intechopen.com/books/open-questions-in-cosmology}{InTech}, 2012), \href{http://cdn.intechopen.com/pdfs/41690.pdf}{ch. 12}.

\bb{iso}M.~Cveti\v{c}, G.W.~Gibbons, D.~Kubiz\v{n}\'{a}k, and C.N.~Pope,
\emph{Black Hole Enthalpy and an Entropy Inequality for the Thermodynamic Volume},
Phys. Rev. D \textbf{84} 024037 (2011). arXiv:1012.2888 [hep-th].

\bibitem{Mann1}D.~Kubiz\v{n}\'{a}k and R.B.~Mann,
\emph{$P-V$ criticality of charged AdS black holes},
JHEP07 033 (2012). arXiv:1205.0559 [hep-th].

\bibitem{Mann2}S.~Gunasekaran, R.B.~Mann, and D.~Kubiz\v{n}\'{a}k,
\emph{Extended phase space thermodynamics for charged and rotating black holes and Born-Infeld vacuum polarization},
JHEP11 110 (2012). arXiv:1205.0559 [hep-th].

\bibitem{AzregE}M.~Azreg-A\"{\i}nou,
\emph{Charged de Sitter-like black holes: quintessence-dependent enthalpy and new extreme solutions},
Eur. Phys. J. C \textbf{75} 34 (2015). arXiv:1410.1737 [gr-qc].

\bibitem{consts}S.P.~Preval \textit{et al.},
\emph{Do the constants of nature couple to strong gravitational fields?},
arXiv:1410.0809 [astro-ph.SR].

\bb{BI}M.~Born and L.~Infeld,
\emph{Foundations of the new field theory},
Proc. Roy. Soc. A \textbf{144} 425 (1934).

\bibitem{irina} I.~Dymnikova and M.~Korpusik,
\emph{Thermodynamics of regular cosmological black boles with the de Sitter interior},
Entropy \textbf{13} 1967  (2011).

\bibitem{IrinaD}I.~Dymnikova,
\emph{Vacuum nonsingular black hole},
Gen. Relativ. Gravit. \textbf{24}, 235 (1992).

\bibitem{irina2} I.~Dymnikova and M.~Korpusik,
\emph{Regular black hole remnants in de Sitter space},
Phys. Lett. B \textbf{685} 12 (2010).

\bibitem{Magli} A.~Burinskii, E.~Elizalde, S.R.~Hildebrandt, and G.~Magli,
\emph{Regular sources of the Kerr-Schild class for rotating and nonrotating black hole solutions},
Phys. Rev. D \textbf{65} 064039 (2002). arXiv:gr-qc/0109085.

\bibitem{Lemos}J.P.S.~Lemos and V.T.~Zanchin,
\emph{Regular black holes: Electrically charged solutions, Reissner-Nordstr\"om outside a de Sitter core},
Phys. Rev. D \textbf{83} 124005 (2011). arXiv:1104.4790 [gr-qc].

\bibitem{conf} M.~Azreg-A\"{\i}nou,
\emph{Regular and conformal regular cores for static and rotating solutions},
Phys. Lett. B \textbf{730} 95 (2014). arXiv:1401.0787 [gr-qc].

\bb{MTZ}C.~Martinez, R.~Troncoso, and J.~Zanelli,
\emph{de Sitter black hole with a conformally coupled scalar field in four dimensions},
Phys. Rev. D \textbf{67} 024008 (2003). arXiv:hep-th/0205319.

\bibitem{Kis}V.V.~Kiselev,
\emph{Quintessence and black holes},
Class. Quantum Grav. \textbf{20} 1187 (2003). arXiv:0210040 [gr-qc].

\bibitem{AAR}M.~Azreg-A\"{\i}nou and M.E.~Rodrigues,
\emph{Thermodynamical, geometrical and Poincar\'e methods for charged black holes in presence of quintessence},
JHEP09 146 (2013). arXiv:1211.5909 [gr-qc].

\bb{on-gtd}M.~Azreg-A\"{\i}nou,
\emph{Geometrothermodynamics: comments, criticisms, and support},
Eur. Phys. J. C \textbf{74} 2930 (2014). arXiv:1311.6963.

\bb{cor1}S.~Carlip,
\emph{Logarithmic corrections to black hole entropy, from the Cardy formula},
Class. Quantum Grav. \textbf{17} 4175 (2000). arXiv:gr-qc/0005017.

\bb{cor2}S.~Das, P.~Majumdar, and R.K.~Bhaduri,
\emph{General logarithmic corrections to black-hole entropy},
Class. Quantum Grav. \textbf{19} 2355 (2002). arXiv:hep-th/0111001.

\bibitem{cor3}A.~Sen,
\emph{Logarithmic corrections to Schwarzschild and other non-extremal black hole entropy in different dimensions},
JHEP04 156 (2013). arXiv:1205.0971 [hep-th].

\bb{cor4}R.~Aros, D.E.~D\'{i}az, and A.~Montecinos,
\emph{Wald entropy of black holes: Logarithmic corrections and trace anomaly},
Phys. Rev. D {\bf 88} 104024 (2013). arXiv:1305.4647 [gr-qc].

\bb{extra1}R.M.~Wald,
\emph{Black hole entropy is Noether charge},
Phys. Rev. D \textbf{48} 3427 (1993). arXiv:gr-qc/9307038.

\bb{extra2}T.~Jacobson, G.~Kang, and R.C.~Myers,
\emph{On black hole entropy},
Phys. Rev. D \textbf{49} 6587 (1994). arXiv:9312023 [gr-qc].

\bb{extra3}V.~Iyer and R.M.~Wald,
\emph{Some properties of Noether charge and a proposal for dynamical black hole entropy},
Phys. Rev. D \textbf{50} 846 (1994). arXiv:9403028 [gr-qc].

\bb{book}H.E.~Stanley, \emph{Introduction to Phase Transitions and Critical
Phemonena} (Oxford University Press, NY, 1987).

\bibitem{AAMR}M.~Azreg-A\"{\i}nou, G.T.~Marques, and M.E.~Rodrigues,
\emph{Phantom black holes and critical phenomena},
JCAP07 036 (2014). arXiv:1405.5745 [gr-qc].

\bibitem{BI1}S.~Fernando and D.~Krug,
\emph{Charged black hole solutions in Einstein-Born-Infeld gravity with a cosmological constant},
Gen. Relativ. Gravit. \textbf{35}, 129 (2003). hep-th/0306120.

\bibitem{BI2} T.K.~Dey,
\emph{Born-Infeld black holes in the presence of a cosmological constant},
Phys. Lett. B \textbf{595} 484 (2004). hep-th/0406169.

\bb{BI3}R.-G.~Cai, D.-W.~Pang, and A.~Wang,
\emph{Born-Infeld black holes in (A)dS spaces},
Phys. Rev. D \textbf{70} 124034 (2004). hep-th/0410158.

\bb{MTZ2}A.-M.~Barlow, D.~Doherty, and E.~Winstanley,
\emph{Thermodynamics of de Sitter black holes with a conformally coupled scalar field},
Phys. Rev. D \textbf{72} 024008 (2005). arXiv:gr-qc/0504087.

\bibitem{reg}E.~Ay\'on-Beato and A.~Garc\'{\i}a,
\emph{Regular black hole in general relativity coupled to nonlinear electrodynamics},
Phys. Rev. Lett. \textbf{80} 5056 (1998). arXiv:gr-qc/9911046.

\bibitem{AzregR}M.~Azreg-A\"{\i}nou,
\emph{Generating rotating regular black hole solutions without complexification},
Phys. Rev. D \textbf{90} 064041 (2014). arXiv:1405.2569 [gr-qc].

\bibitem{para1}K.~Hayashi and T.~Shirafuji,
\emph{New general relativity},
Phys. Rev. D \textbf{19} 3524 (1979) [Addendum-ibid. D \textbf{24} 3312 (1982)].

\bibitem{para2}C.Q.~Geng, C.C.~Lee, and E.N.~Saridakis,
\emph{Observational constraints on teleparallel dark energy},
JCAP01 002 (2012). arXiv:1110.0913 [astro-ph.CO].

\bibitem{para3}C.~Xu, E.N.~Saridakis and G.~Leon,
\emph{Phase-space analysis of teleparallel dark energy},
JCAP07 005 (2012), arXiv:1202.3781 [gr-qc].

\bibitem{Li} G.-Q.~Li,
\emph{Effects of dark energy on P-V criticality of charged AdS black holes},
Phys. Lett. B \textbf{735} 256 (2014). arXiv:1407.0011 [gr-qc].

\bibitem{gerard}G.~Cl\'ement, J.C.~Fabris, and M.E.~Rodrigues,
\emph{Phantom black holes in Einstein-Maxwell-dilaton theory},
Phys. Rev. D \textbf{79} 064021 (2009). arXiv:0901.4543 [hep-th].

\bibitem{thermo} M.E.~Rodrigues and Z.A.A.~Oporto,
\emph{Thermodynamics of phantom black holes in Einstein-Maxwell-dilaton theory},
Phys. Rev. D \textbf{85} 104022 (2012). arXiv:1201.5337 [gr-qc].

\bibitem{van1} A.~Rajagopal, D.~Kubiz\v{n}\'{a}k, and R.B.~Mann,
\emph{Van der Waals black hole},
Phys. Lett. B \textbf{737} 277 (2014). arXiv:1408.1105 [gr-qc].

\bibitem{van2} T.~Delsate and R.B.~Mann,
\emph{Van Der Waals Black Holes in $d$ dimensions},
JHEP02 070 (2015). arXiv:1411.7850 [gr-qc].



\end{thebibliography}
\end{document}